# High-precision measurement of the atomic mass of the electron

S. Sturm[1], F. Köhler[1,2], J. Zatorski[1], A. Wagner[1], Z. Harman[1,3], G. Werth[4], W. Quint[3], C. H. Keitel[1] and K. Blaum[1]

[1] Max-Planck-Institut für Kernphysik, Saupfercheckweg 1, 69117 Heidelberg, Germany

[2] GSI Helmholtzzentrum für Schwerionenforschung, Planckstraße 1, 64291 Darmstadt, Germany

[3] ExtreMe Matter Institute EMMI, Planckstraße 1, 64291 Darmstadt, Germany

[4] Institut für Physik, Johannes Gutenberg-Universität, Staudingerweg 7, 55128 Mainz, Germany

**The quest for the value of the electron's atomic mass has been subject of continuing efforts over the last decades** [1, 2, 3, 4]**. Among the seemingly fundamental constants which parameterize the Standard Model (SM) of physics** [5] **and which are thus responsible for its predictive power, the electron mass $m_e$ plays a prominent role, as it is responsible for the structure and properties of atoms and molecules. This manifests in the close link with other fundamental constants, such as the Rydberg constant $R_\infty$ and the fine-structure constant $\alpha$** [6]**. However, the low mass of the electron considerably complicates its precise determination. In this work we present a substantial improvement by combining a very accurate measurement of the magnetic moment of a single electron bound to a carbon nucleus with a state-of-the-art calculation in the framework of bound-state Quantum Electrodynamics. The achieved precision of the atomic mass of the electron surpasses the current CODATA** [6] **value by a factor of 13. Accordingly, the result presented in this letter lays the foundation for future fundamental physics experiments** [7, 8] **and precision tests of the SM** [9, 10, 11]

Throughout the last decades, the determination of the atomic mass of the electron has been subject to several Penning-trap experiments, as continuing experimental efforts try to further explore the scope of validity of the SM and require an exceedingly precise knowledge of $m_e$. The uniform magnetic field of these traps gives the possibility to compare the cyclotron frequency of the electron with that of another ion of known atomic mass, typically carbon ions or protons. The first such direct determination dates back to 1980, when Gräff *et al.* made use of a Penning trap to compare the cyclotron frequencies of a cloud of electrons with that of protons, which were alternately confined in the same magnetic field, yielding a relative precision of about 0.2 ppm [2]. Since then, a number of experiments have pushed the precision by about 3 orders of magnitude [1, 12, 13, 4]. The latest version of the CODATA compilation of fundamental constants of 2010 lists a relative uncertainty of $4 \cdot 10^{-10}$, resulting from the weighted average of the most precise measurements (Fig. 2). Since the cyclotron frequency of the extremely light electron is subjected to troublesome relativistic mass shifts if not held at the lowest possible energy, direct ultra-high precision mass measurements are particularly delicate. To circumvent this problem, the currently most precise measurements, including this work, pursue an indirect method which allows achieving a previously unprecedented accuracy.

A single electron is bound directly to the reference ion, in this case a bare carbon nucleus (Fig. 1). In this way, it becomes possible to calibrate the magnetic field $B$ at the very place of the electron through a measurement of the cyclotron frequency

$$\nu_c = \frac{1}{2\pi}\frac{q}{m_{ion}}B \tag{1}$$

of the heavy-ion system with mass $m_{ion}$ and charge $q$. While the cyclotron frequency of the strongly bound electron is of no further relevance, the precession frequency of the electron spin, which depends on the electron's magnetic moment $\mu_s$,

$$\nu_L = \frac{2\mu_s B}{h} = \frac{g}{4\pi}\frac{e}{m_e}B, \tag{2}$$

is well defined and reveals information on the mass of the electron $m_e$. A measurement of the ratio of these two frequencies yields $m_e$ in units of the ion's mass:

$$m_e = \frac{g}{2}\frac{e}{q}\frac{\nu_c}{\nu_L}m_{ion} \equiv \frac{g}{2}\frac{e}{q}\frac{1}{\Gamma}m_{ion}, \tag{3}$$

where $\Gamma$ denotes the experimentally determined ratio $\nu_L/\nu_c$. When determining $\Gamma$ of a hydrogen-like carbon ion, which is the defining particle for the atomic mass (apart from the mass and binding energies of the missing electrons, which are sufficiently well known), the remaining unknown in Eq. (3) is the $g$-factor. Advances in Quantum Electrodynamics (QED) theory in recent years allow us to calculate this value with highest precision [14].

In this Letter we present an ultra-precise measurement of the frequency ratio and a state-of-the-art QED calculation for the case of hydrogen-like $^{12}C^{5+}$, which allow to determine $m_e$ with unprecedented accuracy. Exposing the electron to the binding Coulomb field of an atomic nucleus has a profound influence on the $g$-factor. The largest difference from the free-electron case can be deduced from a solution of the Dirac equation in the presence of the Coulomb potential of a nucleus of charge $Z$ and an external, constant and homogeneous magnetic field: $g_{Dirac} = \frac{2}{3} + \frac{4}{3}\sqrt{1-(Z\alpha)^2}$ [15]. This result has to be complemented by various other effects, originating mainly from QED (see Fig. 3). Many of those effects, like the one-loop self-energy and vacuum polarization terms, and the nuclear recoil contribution are known with sufficient numerical accuracy [14, 16, 17]. The main challenge in further improving the theoretical predictions is related to the two-loop (2L) QED effect. This contribution is only known to the first few terms of its expansion in terms of $(Z\alpha)^n ln^k[(Z\alpha)^{-2}]$. The calculation of the expansion coefficients with $n \geq 5$ is beyond the current state of the art, defining the overall theoretical uncertainty. However, we have been able to estimate the uncalculated higher-order contribution $g_{2L}^{(ho)}$ and thus improve on the theoretical value with the help of our recent experimental value $g_{exp}^{Si}$ of the $g$-factor of hydrogen-like silicon ($Z=14$) [18, 19]. This contribution, which dominates the theoretical uncertainty, can be determined from the difference of the experimentally determined $g$-factor and the theoretical prediction, which is the sum of all known terms excluding $g_{2L}^{(ho)}$:

$$g_{2L}^{(ho)}(Z=14) = g_{exp}^{Si} - g_{theo}^{Si} = 2\frac{\nu_L}{\nu_c}\frac{m_e}{m_{ion}}13 - g_{theo}^{Si}. \tag{4}$$

We assume an analytical form of the $Z$-dependence $g_{2L}^{(ho)}(Z)$, and thus obtain an estimate of $g_{2L}^{(ho)}(C)$, presented in Supplementary Table SII. Eq. (4), together with a second formula on the $Z$ dependence on those higher-order terms (see Eq. (S15)), can be solved to yield more accurate values for two variables, namely, the theoretical $g$-factor value for carbon, and the electron mass. The technical details of this calculation and related uncertainty are described in the supplement.

The key tool for our measurements is the Penning trap. The homogeneous magnetic field (in our case 3.7 T), which causes the precession of the spin, also forces the ion on a circular cyclotron motion and in this way confines it in the plane perpendicular to the field

("radial" plane). In order to retain the ion sufficiently long for a precision measurement, we add an electrostatic quadrupole potential, which yields a harmonic motion of the ion along the magnetic field lines ("axial") with frequency $v_z$. Simultaneously, the quadrupole generates two uncoupled harmonic eigenmotions in the radial plane, the modified cyclotron and the magnetron motion, with frequencies $v_+$ and $v_-$, respectively. The trap eigenfrequencies are connected to the free-space cyclotron frequency via the invariance relation $v_c = \sqrt{v_+^2 + v_z^2 + v_-^2}$ [20]. In order to determine these frequencies, a superconducting tank circuit in resonance with the axial motion of the ion serves to transform tiny currents that the ion induces by its oscillation between the trap electrodes into small yet measurable voltage signals.

The interaction with the tank circuit also serves for a weak damping, which brings the ion into thermal equilibrium with the resonator. Electronic noise feedback techniques allow to further cooling, until the effective temperature is well below that of the environment (of 4.2 K) greatly reducing systematic errors. In thermal equilibrium, the ion exactly cancels the thermal noise of the tank circuit, leaving a characteristic "dip" in the detected spectrum (Fig. 4). A fit with a well-known lineshape directly reveals the axial frequency of the ion to sufficient precision. The remaining eigenfrequencies, which do not couple to the resonator directly, have to be detected via mode-coupling to the axial motion. The recent development of the "Pulse and Amplify" (PnA) technique [21] has enabled us to perform phase-sensitive measurements of the modified cyclotron frequency at lowest energies below the detection threshold of the image-current amplifier, a significant improvement compared to the established 'pulse and probe' technique [22]. Combined with the axial and magnetron frequency information from dip fits, the invariance relation allows to calculate the free-space cyclotron frequency, which is a measure of the magnetic field at the ion's location.

The Larmor precession frequency, nominally 105 GHz in our case, cannot be detected directly with the image current detector. Instead, the Zeeman splitting of the bound electron's spin is probed with a microwave excitation. The key requirement for this is the ability to detect the spin state with the continuous Stern-Gerlach effect [23]. To this end, a strong magnetic field inhomogeneity, is generated by an electrode made from ferromagnetic material. In our setup the quadratic portion of this bottle-shaped field amounts to $B_2=10^4$ T/m$^2$ [24]. In this inhomogeneous field, the magnetic moment couples to the axial motion and causes a small, spin-dependent frequency difference. Provided all other influences on the axial frequency, notably the ion's energy and the voltages applied to the trap, can be sufficiently well controlled, the determination of the axial frequency of the ion becomes a quantum non-demolition measurement of the electron's spin. We use a double-trap setup to spatially separate the spin analysis in the inhomogeneous field of the "Analysis Trap" (AT) and the high-precision eigenfrequency measurement in the "Precision Trap" (PT) (Fig. 4). During the experiment, the ion is adiabatically shuttled between these two traps. After determining the initial spin-state in the AT, the ion is transported to the PT, where a microwave excitation at a random frequency offset with respect to the expected Larmor frequency probes the Zeeman splitting in coincidence with the same time as the 'pulse and amplify' measurement of the cyclotron frequency, which suppresses fluctuations of the magnetic field. The axial frequency, which is basically independent of the magnetic field, is measured before and after the 'pulse and amplify' cycle and interpolated. After transporting the ion back to the AT, an analysis of the spin state allows us to detect a possible successful spin-flip in the PT. By repeating this process (see Fig. 1) several hundred times it becomes possible to map the probability of spin-flips in the homogeneous magnetic field of the PT as a function of the frequency ratio Γ (rightmost panel of Fig. 1).

The dominant systematic uncertainty arises from the self-interaction mediated by image charges and currents in the trap electrodes. In contrast to the free electron case [9], the retardation of the field and the resultant damping through a coupling to modes of the trap acting as a cavity is negligible owing to the very much higher cyclotron wavelength. However, instead, the influence of the immediate Coulomb interaction - that is, image charges - is enhanced. Even though the resultant shift can be readily calculated, finite machining accuracies and the imperfect knowledge of the ion's geometric position impose a relative uncertainty of $\delta v_c/v_c = 1.5 \cdot 10^{-11}$.

The extrapolated frequency ratio $\Gamma_0' \cong \Gamma(E_+ = 0)$, corrected for all systematic shifts (Table 1), yields the final value $\Gamma_0 = 4376.210\ 500\ 89\ (11)(7)$, with the statistical and systematic uncertainties, respectively, given in parentheses. The theoretical prediction of the $g$-factor presented here (see Table SII of the supplement) allows calculating the mass of the electron in units of the ion's mass. By correcting for the mass of the missing electrons and their respective atomic binding energies, taken from [25], we can finally calculate $m_e$ in atomic mass units:

$$m_e = 0.000\ 548\ 579\ 909\ 067(14)(9)(2)\ . \tag{5}$$

The first two errors are the statistical and systematic uncertainties of the measurement, while the third value represents the uncertainties of the theoretical prediction of the $g$-factor and the electron binding energies. The theoretical result for the $g$-factor, with corrections obtained from the experimentally determined value for hydrogen-like $^{28}$Si$^{13+}$ [18], implicitly assumes the correctness of QED. However, the so far untested higher-order contribution determined in this work scales with $(Z\alpha)^5$ and thus contributes less than $10^{-11}$ in relative terms for the $^{12}$C$^{5+}$ system.

The relative precision of $3 \cdot 10^{-11}$ for $m_e$ obtained in this work surpasses the current CODATA [6] averaged literature value by a factor of 13 and the previous best measurement [3] by a factor of 17 (see Fig. 2). Furthermore, our result allows giving the electron-proton mass ratio with a relative precision of 94 ppt, solely determined by the proton mass value:

$$\frac{m_p}{m_e} = 1836.152\ 673\ 77\ (17). \tag{6}$$

The main limitations seen in this work are the uncertainty resulting from the ion's self-interaction with its own image-charge in the trap electrodes and the temperature of the ion in connection with the temporal stability of the magnetic field.

Our result sets the stage for future ultra-high precision tests of the Standard Model at low energies. One example is the determination of the fine-structure constant $\alpha$ via a measurement of the recoil momentum exerted on an atom upon absorption of a photon [26]. The electron atomic mass presented in this letter, combined with the Rydberg constant [6], the atomic mass of Rubidium [27] and an atom interferometric measurement of $h/m_{\text{Rb}}$ [26], yields a value for $\alpha$. By inserting this value into the Kinoshita theory for the $g$-factor of the free electron, it is possible to test the Standard Model through the Gabrielse experiment and, among others, to observe the unification of the electro-weak interaction at low energies and probe the existence of light dark-matter particles [11]. Furthermore, the new value for $m_e$ paves the way to probe the validity of QED at highest field strength through $g$-factor determinations in heavy, highly charged ions [18, 7, 8].

**Supplementary Information** is linked to the online version of the paper at www.nature.com/nature.

**Acknowledgements:**

This work was supported by the Max-Planck Society, the EU (ERC Grant No. 290870 - MEFUCO), the IMPRS-QD, GSI and the Helmholtz Alliance HA216/EMMI.

**Author Contributions:** SS, FK and AW performed the experiment. SS and FK performed the data analysis. JZ performed the QED calculations. SS, FK, KB, JZ and ZH prepared the manuscript. SS and FK prepared the experimental supplement. JZ and ZH prepared the theory supplement. All authors discussed the results and contributed to the manuscript at all stages.

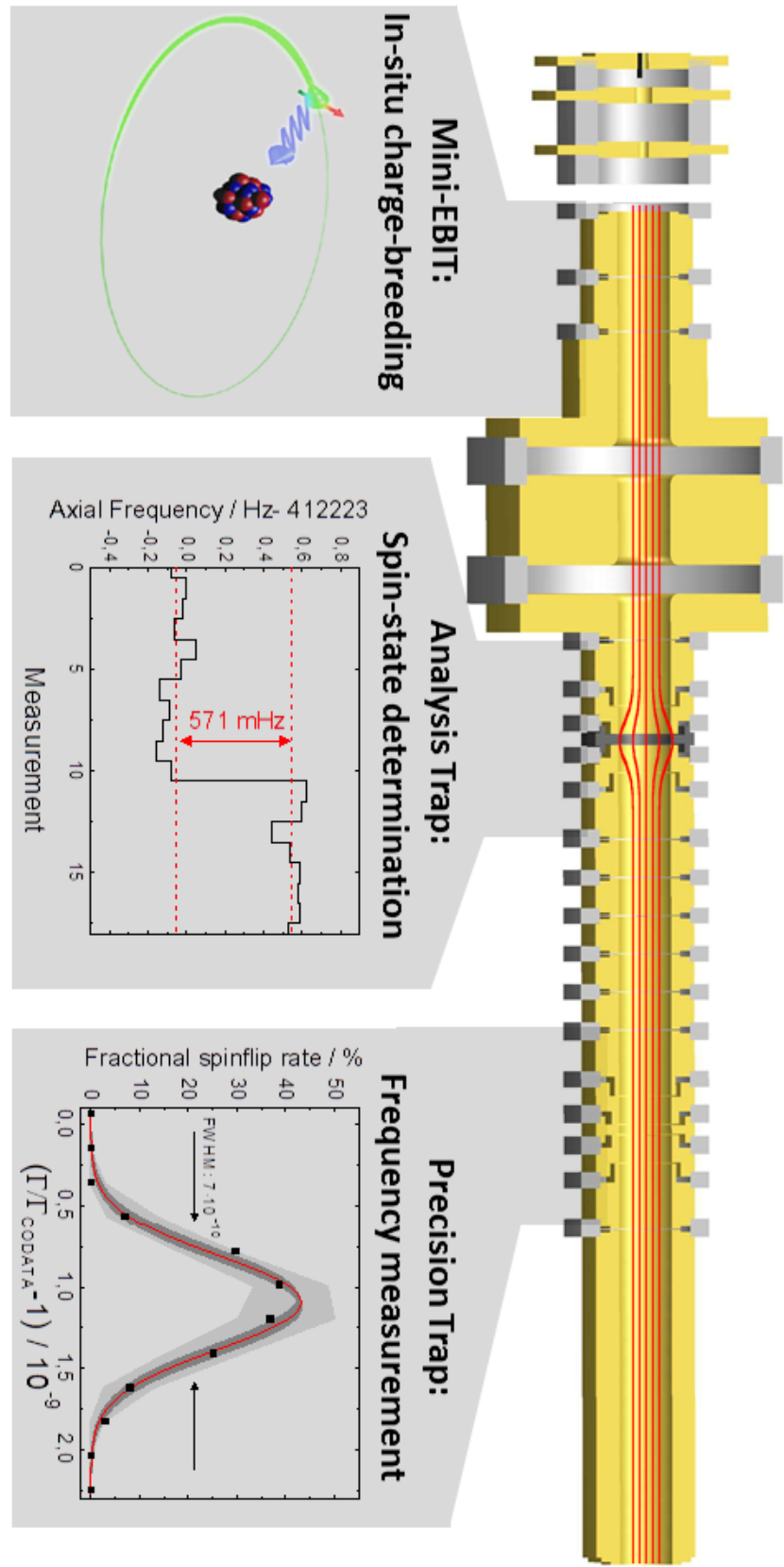

**Figure 1 | The triple Penning-trap setup used in this work.** Yellow: electrodes; grey: insulation rings; red: magnetic field lines (sketch). Highly charged ions can be charge-bred *in situ* inside the hermetically closed cryogenic vacuum, allowing for virtually infinite measurement time (lower panels). The magnetic bottle in the Analysis Trap, used for the spin-state detection, is spatially separated from the very homogeneous field in the Precision Trap, which allows precise measurements of the ion's eigenfrequencies. For details see text

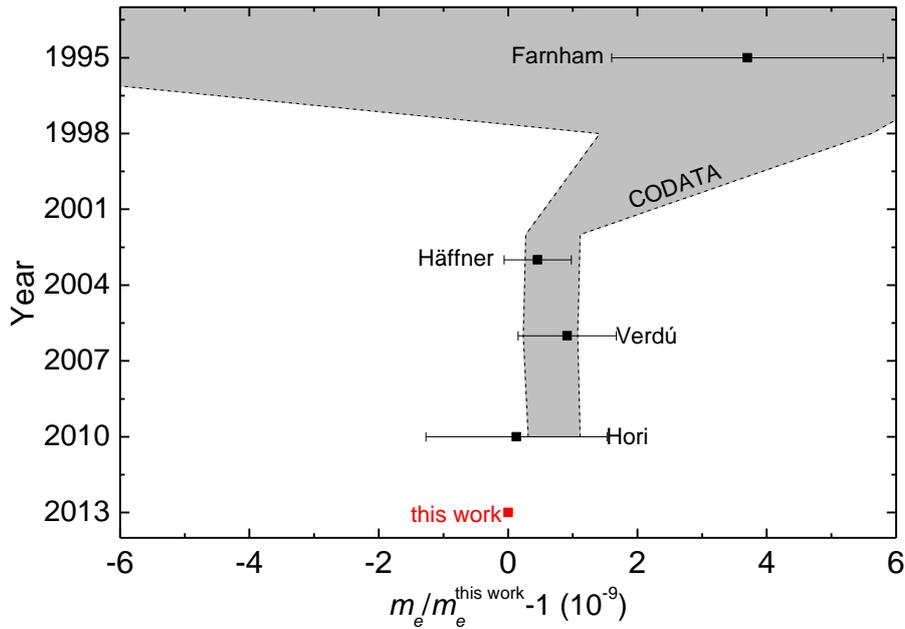

**Figure 2 | History of electron mass measurements.** The last direct cyclotron frequency determination dates back to 1995, the more recent values are all indirect determinations based on QED predictions of *g*-factors or transition energies. The gray band is the 1σ-confidence interval of the CODATA evaluations of the respective years. Recently, a flaw in the handling of systematic shifts in one of the input values was found, suggesting a world average value about 0.5σ smaller than the current CODATA value. For details see the supplement. The error bar of the new value is hidden by the square symbol.

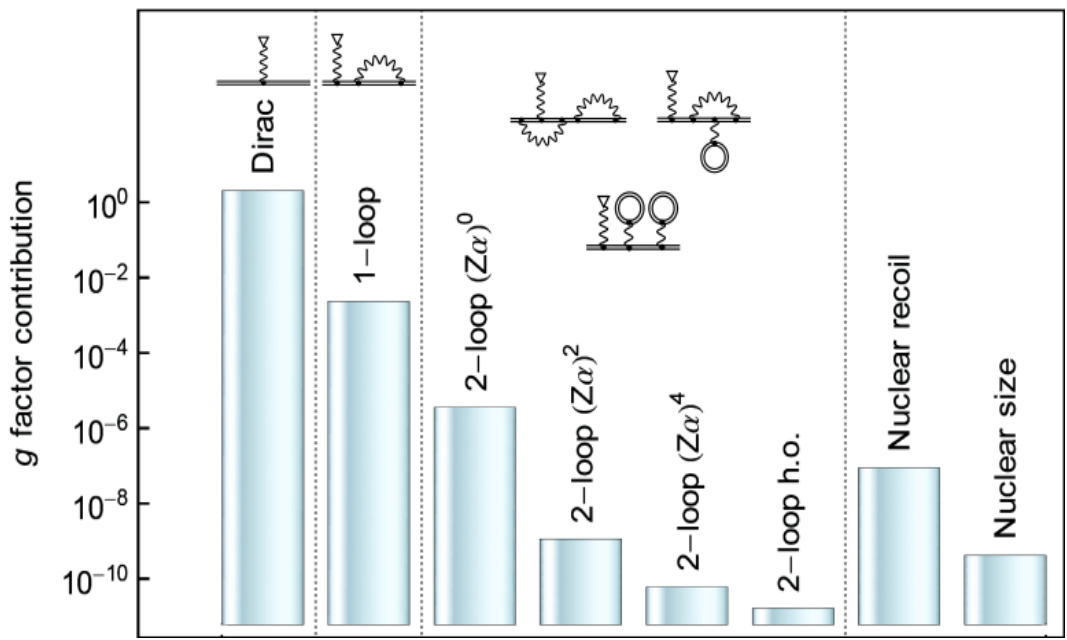

**Figure 3 | The magnitude of the relevant theoretical contributions to the bound electron *g*-factor in $^{12}C^{5+}$:** The leading Dirac contribution, one- and two-loop BS-QED corrections, and nuclear effects (see also the supplement). Some representative Feynman diagrams [16] corresponding to the QED terms are shown.

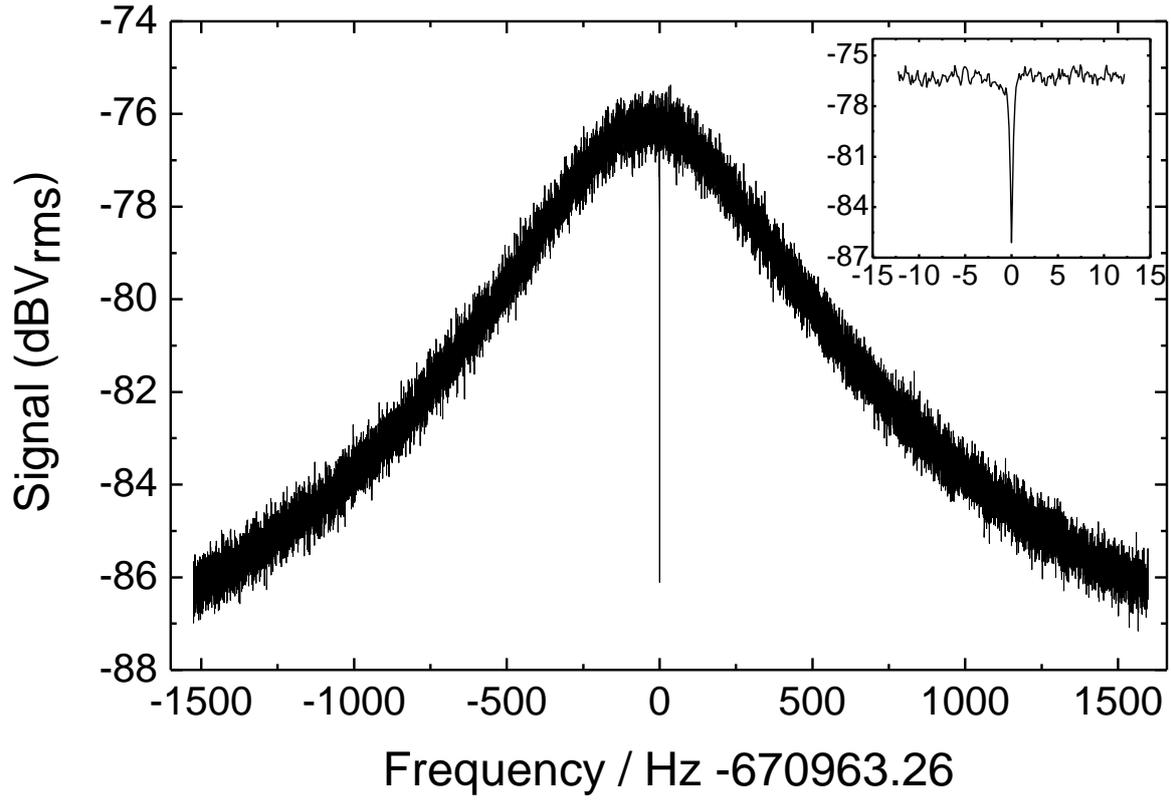

**Figure 4 | Axial dip signal of a single $^{12}C^{5+}$ ion**, used to determine the axial frequency. The linewidth is only about 0.4 Hz. The inset shows a close-up of the dip feature.

| Effect | Correction (ppt) | Uncertainty (ppt) |
|---|---|---|
| Image charge | -282.4 | 14.1 |
| Image current | 2.2 | 0.5 |
| Residual electrostatic Anharmonicity | 0 | 0.25 |
| Axial & magnetron Temperature | 0.04 | 0.04 |
| Ionic mass $^{12}C^{5+}$ | - | 0.1 |

**Table 1 | Relative systematic corrections applied to the measured frequency ratio and the corresponding uncertainties.** The small shift due to the residual cyclotron energy is eliminated by an extrapolation of the frequency ratios measured at different energies. For details see supplementary material.